\documentclass[aps,prx, twocolumn,superscriptaddress]{revtex4-1}

\usepackage{txfonts}
\usepackage{hyperref}
\usepackage{graphicx}
\usepackage{color}
\usepackage{lineno}

\hypersetup{colorlinks,
linkcolor=blue,
filecolor=blue,
urlcolor=blue,
citecolor=blue}
\begin{document}
\title{In-plane thermal diffusivity determination using beam-offset frequency-domain thermoreflectance with a one-dimensional optical heat source}

\author{Kai Xu}
\affiliation{Institut de Ci\`encia de Materials de Barcelona, ICMAB-CSIC, Campus UAB, 08193 Bellaterra, Spain}

\author{Jiali Guo}
\affiliation{Institut de Ci\`encia de Materials de Barcelona, ICMAB-CSIC, Campus UAB, 08193 Bellaterra, Spain}
\author{Grazia Raciti}
\affiliation{Physics Department, University of Basel, Klingelbergstrasse 82, CH-4056 Basel, Switzerland}

\author{Alejandro R. Goni}
\affiliation{Institut de Ci\`encia de Materials de Barcelona, ICMAB-CSIC, Campus UAB, 08193 Bellaterra, Spain}
\affiliation{ICREA, Passeig Llu\'{\i}s Companys 23, 08010 Barcelona, Spain}

\author{M. Isabel Alonso}
\affiliation{Institut de Ci\`encia de Materials de Barcelona, ICMAB-CSIC, Campus UAB, 08193 Bellaterra, Spain}

\author{Xavier Borrisé}
\affiliation{Catalan Institute of Nanoscience and Nanotechnology (ICN2), CSIC and the Barcelona Institute of Science and Technology (BIST), Building ICN2, Campus UAB, E-08193 Bellaterra, Barcelona, Spain}

\author{Ilaria Zardo}
\affiliation{Physics Department, University of Basel, Klingelbergstrasse 82, CH-4056 Basel, Switzerland}

\author{Mariano Campoy-Quiles}
\affiliation{Institut de Ci\`encia de Materials de Barcelona, ICMAB-CSIC, Campus UAB, 08193 Bellaterra, Spain}

\author{Juan Sebastián Reparaz}
\email{jsreparaz@icmab.es}
\affiliation{Institut de Ci\`encia de Materials de Barcelona, ICMAB-CSIC, Campus UAB, 08193 Bellaterra, Spain}

\begin{abstract}
We present an innovative contactless method suitable to study in-plane thermal transport based on beam-offset frequency-domain thermoreflectance using a one-dimensional heat source with uniform power distribution. Using a one-dimensional heat source provides a number of advantages as compared to point-like heat sources, as typically used in time- and frequency-domain thermoreflectance experiments, just to name a few: (i) it leads to a slower spatial decay of the temperature field in the direction perpendicular to the line-shaped heat source, allowing to probe the temperature field at larger distances from the heater, hence, enhancing the sensitivity to in-plane thermal transport; (ii) the  frequency range of interest is typically $<$ 100 kHz. This rather low frequency range is convenient regarding the cost of the required excitation laser system but, most importantly, it allows the study of materials without the presence of a metallic transducer with almost no influence of the finite optical penetration depth of the pump and probe beams on the thermal phase lag, which arises from the large thermal penetration depth imposed by the used frequency range.   
We also show that for the case of a harmonic thermal excitation source, the phase lag between the thermal excitation and thermal response of the sample exhibits a linear dependence with their spatial offset, where the slope is proportional to the inverse of the thermal diffusivity of the material. We demonstrate the applicability of this method to the cases of: (i) suspended thin films of Si and PDPP4T, (ii) Bi bulk samples, and (iii) Si, glass, and highly-oriented pyrollitic graphite (HOPG) bulk samples with a thin metallic transducer.
Finally, we also show that it is possible to study in-plane heat transport on substrates with rather low thermal diffusivity, e.g., glass, even using a metallic transducer. We achieve this by an original approach based on patterning the transducer using focused ion beam, with the key purpose of limiting in-plane heat transport through the thin metallic transducer.
\end{abstract}

\maketitle

\section{Introduction}
The study of the thermal conductivity (or diffusivity) tensor ($\kappa_{ij}$) in bulk and low dimensional materials has gained considerable momentum in recent years. For example, in layered materials where the in-plane and out-of-plane components of the thermal conductivity exhibit strong anisotropy, or for the case of the ``artificial anisotropy" induced by low dimensionality in the case of crystalline (or semi-crystalline) thin films. A large number of experimental methods to study the out-of-plane components of the thermal conductivity have been developed and successfully demonstrated using different methodologies, e.g., based on electrical or optical methods.\cite{Ly2021,Tong2006,Ramu2012,Mishra2015a,Schmidt2008,Feser2012,Feser2014,Rodin2017,Jiang2017,Jiang2018,Li2018,Rahman2018,Yuan2019,Qian2020,Tang2021,Cahill1990,Lu2001,Cahill2004,perez2022} On the other hand, the study of in-plane thermal transport is comparatively more challenging due to the lack of sensitivity to this component of most methods, among other reasons.
%
In fact, most experimental approaches to study thermal anisotropy are based in achieving experimental sensitivity to in-plane thermal anisotropy, which has been demonstrated through different experimental configurations by several research groups.\cite{Tong2006,Schmidt2008,Feser2012,Feser2014,Wilson2014c,Medvedev2015,Lee2015,Handwerg2015,Handwerg2016a,Rodin2017,Kwon2017,Slomski2017,Jiang2018,Cheng2019,Tang2021,Pang2021}
Electrical methods such as, e.g., the recent extension of the 3-Omega method\cite{Handwerg2015,Handwerg2016a},
where the heater and thermometer are located in different spatial position on the surface of the sample, have been proven successful to study thermally anisotropic materials, however, these approaches result most convenient for electrically insulating samples, otherwise, the metallic transducer must be electrically insulated from the sample, a task that in some cases can be challenging due to the presence of current leakage from the metallic transducer to the sample. Additionally, electrical methods also require considerable fabrication efforts since a specific sample must be fabricated for each in-plane direction that aims to be investigated.
Alternatively, several contactless methods typically based on time- or frequency-domain thermoreflectance, have also been successfully demonstrated to address thermal anisotropy.\cite{Li2018,jiang1,jiang2,Qian2020,Ziade2020,perez2022,jiang2}
Three main experimental strategies were followed: (i) beam-offset experiments based on time- or frequency domain thermoreflectance using a focused Gaussian (0D, isotropic) heat source\cite{Rodin2017,Qian2020,Tang2021,jiang2},
(ii) collinear beam experiments using an elliptically-shaped (2D, anisotropic) heat source,\cite{jiang2,Li2018}
and (iii) collinear beam experiments using a line-shaped (1D, anisotropic) heat source\cite{perez2022} (resembling the geometry used in the 3-Omega method but in a contactless fashion). Although all these methodologies have been proven successful to determine the complete elements of the thermal conductivity tensor, they present some aspects which have the potential to be improved as we demonstrate through the novel approach developed in this work. For example, the methods developed within (i) and (ii) typically suffer from the influence of the shape of the heat source (actually its spatial energy distribution) on the acquired data, which originates from the rather small spatial offsets that can be set between the heat source and the probe. Typically, for 0D heat sources (approximated as focused Gaussian beams) large offsets cannot be achieved due to the rapid spatial decay of the thermal field. On the other hand, the method developed in (iii) avoids the explicit knowledge of the shape of the heat source, similarly to the case of the 3-Omega method,\cite{Cahill1990} but at the expense of precisely knowing the temperature coefficient of reflectance of the surface of the sample (or transducer in most cases), since the absolute temperature must be known to compute the thermal conductivity.

Here, we demonstrate an original experimental approach with enhanced sensitivity to in-plane heat transport, which is based on using a 1D heat source with uniform power distribution along its long axis, but a point-like probe spot. We show that the 1D geometry of the heat source leads to a slower spatial decay of the temperature field as compared to 0D heat sources, hence, allowing to probe the temperature field at relatively large spatial offsets from the heat source. The present approach is based on measuring the phase lag between the thermal excitation and the detection, hence, rendering the thermal diffusivity of the studied samples. 
%
In addition, one of its key advantages is that, for harmonic excitation sources, the phase lag ($\phi$) between the thermal excitation (line-shaped heater) and the detection exhibits a linear relation with their in-plane spatial offset ($\Delta x$), where the slope, $\partial \phi(f) /\partial \Delta x$, is proportional to the thermal diffusivity of the sample for a given modulation frequency, $f$, of the heater. The linear relation between the phase lag and the spatial offset considerably simplifies the data analysis process, i.e., the thermal diffusivity (or thermal conductivity) of the samples is readily obtained through a linear fit of $\phi(\Delta x,f)$. Furthermore, the one-dimensional character of the heat source sets the  frequency range of interest to $f <$ 100 kHz, which  allows the study of materials without the presence of a metallic transducer with almost no influence of the finite optical penetration depth of the pump and probe beams on the thermal phase lag.\cite{Wang2016}

We apply this method to study the in-plane thermal diffusivity of 2D and 3D materials, which in combination with the 1D heat source leads to different heat flow geometries. In particular we studied: (i) suspended Si and PDPP4T (poly(3-([2,20:50,20’-terthiophen]-5-yl)-2,5-bis(2-decyltetradecyl)-6-(thiophen-2-yl)-2,5-dihydropyrrolo[3,4-c]pyrrole-1,4-dione)) thin films with different thicknesses (2D sample; 1D heat flow), (ii) Bi bulk sample without metallic transducer (3D sample; 2D heat flow), and (iii) Si and glass substrates with a 60 nm thick Au transducer (3D sample; 2D heat flow). Finally, we generalize the method for situations where the use of a metallic transducer is unavoidable showing that it is possible to control (and avoid) in-plane heat flow within the transducer by patterning, hence, enhancing the sensitivity to the in-plane thermal diffusivity of the sample independently of the presence of the metallic transducer.

\section{Solution of the heat equation for 1D heat sources} \label{section2}
\begin{figure}[t]
 \includegraphics[scale=0.58,angle=0]{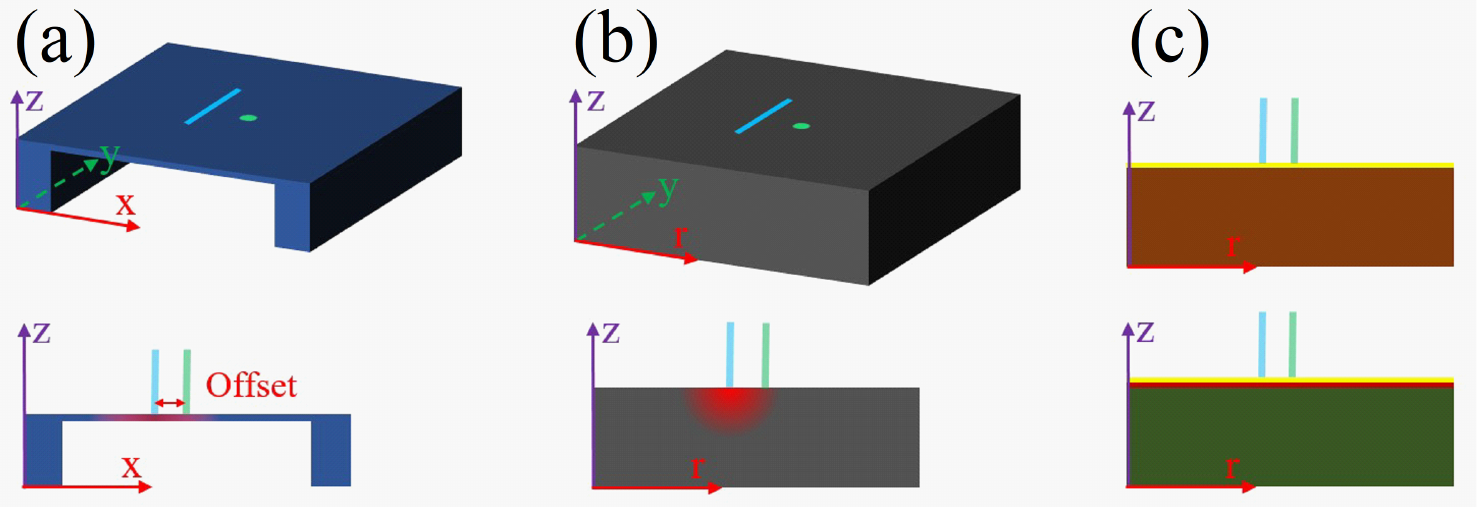}
 \caption{Schematics of the different heat transport geometries studied. In all cases the heater has a line-shaped geometry, whereas the probe beam is given by a diffraction limited focused Gaussian spot. (a) A quasi-2D sample consisting on a suspended thin film. (b) Bulk sample with no transducer. (c) Bulk sample with a transducer or thin film deposited on its surface.}
 \label{fig1}
 \end{figure}

In this section we provide the solution of the heat diffusion equation (parabolic approximation) for the  case of 1D heat source with uniform power distribution along its principal axis, for the case of 2D and 3D materials, and under harmonic excitation conditions. We recall that the key advantage of using a 1D heat source is that no heat propagation occurs along the direction parallel to the line-shaped heat source, hence, reducing the dimensionality of the heat flow geometry and, consequently, of the corresponding heat diffusion equation. Figure \ref{fig1} displays a schematics of each of the heat flow geometries we studied. 


\subsection{1D HEAT FLOW: 1D Heat Source \& 2D Sample}
We start by analyzing the case of a 1D heat source with uniform power distribution along its long axis applied to a quasi-2D material such as suspended membranes. This case is well described by a 1D heat flow geometry, i.e., no heat flow is present along the direction parallel to the line-shaped heat source. The differential equation which describes the system response is simply the 1D heat diffusion equation as follows: 

\begin{equation}
D\frac{\partial^2 T}{\partial x^2}-\frac{\partial T}{\partial t}=-\frac{Q_V}{\rho C_p} \label{eq1}
\end{equation}

\noindent where $T=T(x,t)$ is the temperature,  $t$ and $x$ are the temporal and spatial coordinates, respectively, and $x$ is perpendicular to the line-shaped heater, $D$ is the thermal diffusivity, $\rho$ is the density of the material, $C_p$ is the specific heat capacity, and $Q_V$ is the volumetric heat generation rate. We follow the solution by Salazar,\cite{Salazar2006} where Eq. \ref{eq1} is solved under harmonic excitation conditions, $p(t)=p_0[1+\cos(\omega t)]/2=\Re[p_0(1+e^{i\omega t})/2]$, where $p(t)$ is the time-dependent injected power. In short, the time dependent solution for $T(x,t)$ is obtained by separation of variables as: $T_{ac}(x,t)=\Re[\theta(x)e^{i\omega t}]$. Solving the spatial part of Eq. \ref{eq1} for $\theta(x)$ and multiplying by $e^{i\omega t}$ renders the complete solution:

\begin{equation}
 T(x,t)=\Re\left[\frac{p_0}{2\kappa q}e^{-qx}e^{i\omega t}\right]
\label{eq2}
 \end{equation}

or,

\begin{equation}
 T(x,t)=\frac{\sqrt{2}p_0 \mu_\omega}{4\kappa}e^{-x/\mu_\omega}\cos{\left(\frac{x}{\mu_\omega}-\omega t+\frac{\pi}{4}\right)}
\label{eq3}
 \end{equation}

\noindent where $p_0$ is the maximum amplitude of the absorbed power, $\omega$ is the angular frequency, $\kappa$ is the thermal conductivity, $D$ is the thermal diffusivity, $\mu=\sqrt{2D/\omega}$ is the thermal penetration depth, and $q=\sqrt{i\omega/D}$. Equation \ref{eq3} results particularly interesting if considering the spatial dependence of its argument, whose time independent component corresponds to the phase lag measured in a frequency-domain thermoreflectance (FDTR) experiment. Besides the time dependent component ($\omega t$) which averages to zero after each cycle, the phase lag, $\phi=x/\mu_\omega$, exhibits a linear dependence on the spatial coordinate, $x$. Experimentally, $\phi(x)$ is evidenced as the spatial offset between the line-shaped heat source (1D) and the probe beam (0D) positions. Taking the derivative of the phase lag, $\phi$, with respect to the spatial coordinate, $x$, renders:

\begin{equation}
 \frac{\partial{\phi}}{\partial{x}} = \mu_\omega^{-1}=\sqrt{\frac{\omega}{2D}}=\sqrt{\frac{\pi }{D}}f^{1/2}
\end{equation}

\noindent where $\omega=2\pi f$, and $f$ is the excitation frequency in Hz. Taking the second derivative with respect to $f^{1/2}$ of the previous equation results in: 

\begin{equation}
 \frac{\partial^2{\phi}}{(\partial{x})(\partial{f^{1/2})}} = \sqrt{\frac{\pi }{D}}
 \label{eqderivative}
\end{equation}

\noindent which renders the thermal diffusivity of the specimen. We remark that taking the second order derivative of the phase lag, $\phi$, with respect to $x$ and $f^{1/2}$ is a convenient approach since it avoids the explicit dependence on the absolute calibration of the spatial and frequency scales. This is particularly relevant for the spatial coordinate ($x$), which is prone to larger determination errors as compared to the thermal excitation frequency ($f$). 

\subsection{2D HEAT FLOW: 1D Heat Source and 3D Sample} \label{2Dheatflowbulkwotransducer}

In this section, we study the case of a 1D heat source with uniform power distribution along its long axis applied to a 3D semi-infinite isotropic material. This case is well described by a 2D heat flow geometry, since no heat flow is present along the direction parallel to the line-shaped heat source, similarly as in the previous section. The differential equation which describes the system response is  the 2D heat diffusion equation, which we write in cylindrical coordinates given the symmetry imposed by the 1D heat source as follows: 

\begin{equation}
\frac{1}{r}\frac{\partial}{\partial r}\left(D \cdot r\frac{\partial T}{\partial r}\right)-\frac{\partial T}{\partial t}=-\frac{Q_V}{\rho C_p}
\label{eq5}
\end{equation}

\begin{figure}[t]
 \includegraphics[scale=0.6,angle=0]{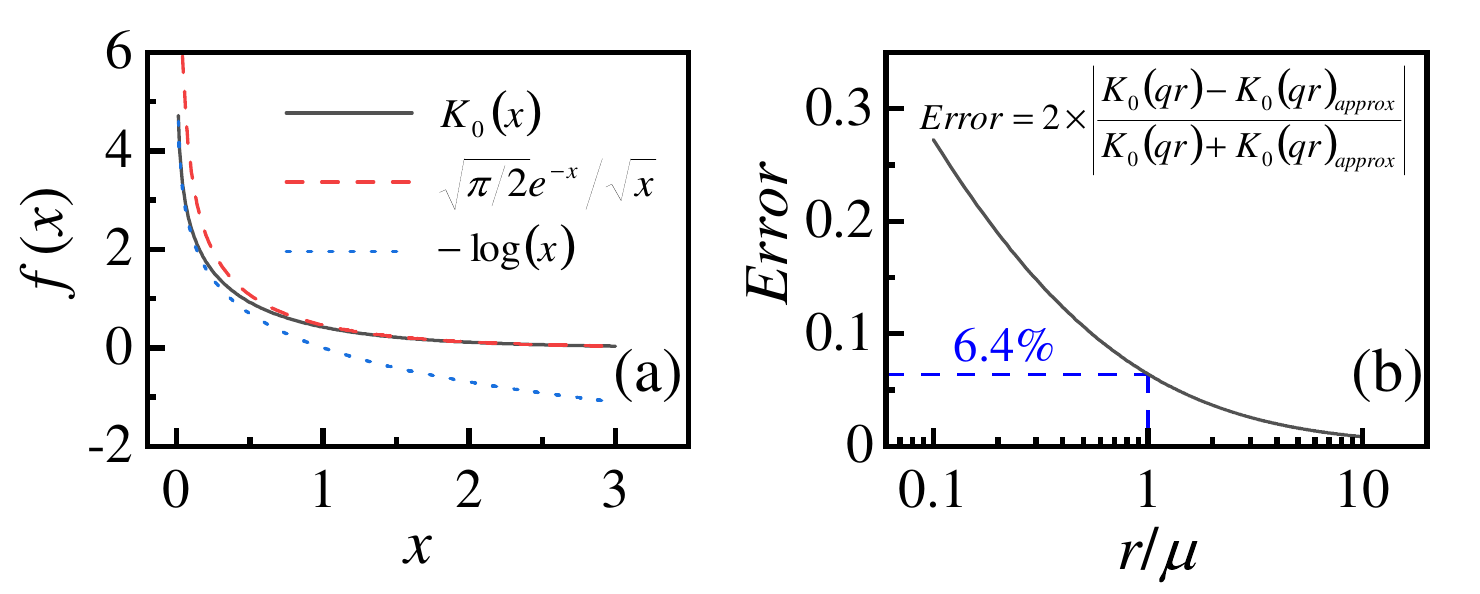}
 \caption{The left panel displays the different approximations to the modified zeroth order Bessel function of the second kind, $K_0(x)$. The right panel shows the error introduced by the approximation $K_0(qr)\approx \sqrt{\pi/2}e^{-qr}/\sqrt{qr}$, which is valid for $|qr|\gg 1$, which is the limit where the thermal penetration depth is smaller than the spatial coordinate.}
 \label{figK0}
 \end{figure}

The solution for $T(r,t)$ is similar to that originally obtained for the case of the 3-Omega method,\cite{Cahill1990} and its optical analogue, Anisotropic Thermoreflectance Thermometry\cite{perez2022} (ATT). Note that due to the radial symmetry, we use the radial coordinate, $r$, instead of the Cartesian coordinate, $x$. In other words, the heat flow geometry is 2D, and the solution for the ideal case of an uniform and infinitely long and narrow line-shaped heater is given by the modified zeroth-order Bessel function of the second kind, $K_0(qr)$.\cite{Carslaw1986, Cahill1990} Similarly as we have proceeded in the previous section, a harmonic thermal excitation source leads to the expression for the time-dependent temperature oscillations, i.e., after multiplication by $e^{i\omega t}$ :

\begin{equation}
 T(r,t)=\Re{[K_0(qr)e^{i \omega t}]}
\end{equation}

\noindent where we recall that $q=\sqrt{i\omega/D}$, $|q|=\sqrt{\omega/D}$, $i^{-1/2}=[1-i]/\sqrt{2}$, $i^{1/2}=[1+i]/\sqrt{2}$, and $i^{1/4}=-i$. It is particularly interesting to study the limiting cases with $|qr|\ll 1$, and $|qr|\gg 1$. The case with $|qr|\ll 1$ ($|r|\ll \sqrt{2}\mu$), leads to the solution obtained for the case of the 3-Omega and ATT methods.\cite{Cahill1990,perez2022} Here, the thermal penetration depth ($\mu$) is much larger than the typical values of radial coordinate, $r$. Note that in the previous example the pump (heater) and probe (thermometer) beams are collinear.
For $|qr|\rightarrow 0$, then $K_0(qr) \propto -ln(qr)$, which is the approximation used to obtain the thermal conductivity, e.g., through the 3-Omega method.\cite{Cahill1990}
On the other hand, in the limit where $|qr|\gg 1$, it can be shown\cite{Caruso2021} that $K_0(qr)\approx \sqrt{\pi/2}e^{-qr}/\sqrt{qr}$. Given the nature of our experiments, we are particularly interested in the limit with $|qr|\gg 1$, which originates in the large offset ($r=\Delta x$) imposed between the 1D heat source and the probe spot as shown in Fig. \ref{fig1}. Figure \ref{figK0}(left) displays the approximations of $K_0(qr)$ for the limiting cases with $|qr|\gg 1$, and $|qr|\ll 1$. In Fig.\ref{figK0}(right) we show the error introduced by the approximation  $K_0(qr)\approx \sqrt{\pi/2}e^{-qr}/\sqrt{qr}$, in terms of the ratio between the spatial coordinate ($r$) and the thermal penetration depth ($\mu$). We consider this approximation to be valid for $r > \mu$, since the error introduced is similar to the experimental error ($\approx$ 5\%)
Hence, the expression for the temperature oscillations is obtained as follows:

\begin{equation}
 T(r,t)=\frac{p_0}{\pi \kappa l}\Re {[K_0(qr) e^{i \omega t}]}\approx \frac{p_0}{\pi \kappa l}\Re \left[ \sqrt{\frac{\pi}{2qr}}e^{-qr}e^{i\omega t}\right ]
\end{equation}

\noindent where $l$ is the length of the line-shaped heat source. The former expression can be rearranged after taking the real part of the argument as follows:

\begin{equation}
 T(r,t)\approx \frac{p_0}{\sqrt[4]{2}\kappa l }\sqrt{\frac{\mu_\omega}{2\pi r}}e^{-r/\mu}\cos\left(\frac{r}{\mu_\omega}-\omega t+\frac{\pi}{8}\right)
\label{T2Dheatflow}
\end{equation}

Taking the second order derivative of the argument of the previous expression ($\phi'=r/\mu_\omega-\omega t+\pi/8$) with respect to $r$ and $f^{1/2}$, we obtain:

\begin{equation}
 \frac{\partial^2{\phi'}}{(\partial{r})(\partial{f^{1/2})}} = \sqrt{\frac{\pi }{D}}
 \label{eqderivative2}
\end{equation}

\noindent which is the same expression that we have obtained for the case of 1D heat flow in the previous section (see Eq. \ref{eqderivative}). The origin of such similarity between the 1D and 2D heat flow cases, is the exponential limit of $K_0(qr)$ for $|qr|\gg 1$. We note that the previous expression is not valid for $|qr|\ll 1$, since in this case the spatial dependence of the temperature field is $T\propto K_0(qr) \propto -ln(qr)$.

\subsection{2D HEAT FLOW: 1D Heat Source and 3D multilayer}

The solution for the case of a 3D multilayer sample can be obtained using the 
approach developed by Borca-Tasciuc,\cite{Borca-Tasciuc2001} as well as the derivation for line-shaped heat sources in Refs. \onlinecite{cahill2004, perez2022}. Here, we only provide the expression that must be numerically solved to obtain the solution for the temperature oscillations. The solution is obtained considering that the line-shaped heater has Gaussian power distribution in direction perpendicular to its long axis, and that the focused probe beam is considered as a Delta function of the spatial coordinate, $\delta(x-x_0)$, where $x_0$ is the absolute position of the line-shaped heater. The previous approximation is valid when the diameter of the focused probe beam is diffraction limited, $d_{probe}<1$ $\mu$m. Using the same notation as Borca-Tasciuc, the frequency-dependent temperature oscillation for different offsets, $\Delta x =x-x_0$, between the heat source and the probe spot positions are: 
 
 \begin{equation}
 T(\Delta x,\omega) =\frac{p_0}{\pi\kappa^{\perp}_1 l}\int_{0}^{\infty}\frac{\exp[-\sigma_{p}^2 \xi^2/8]\cos(\xi\Delta x)}{A_1(\xi)B_1(\xi)} d\xi \label{solution_multilayers}
 \end{equation}
 
 \noindent where $A_1$ and $B_1$ are defined as:
 
 \begin{eqnarray}
 A_{j-1}=\frac{A_j\frac{\kappa^\perp_j B_j}{\kappa^\perp_{j-1}B_{j-1}}-\tanh(B_{j-1}d_{j-1})}{1-A_j\frac{\kappa^\perp_{j}B_j}{\kappa^\perp_{j-1}B_{j-1}}\tanh(B_{j-1}d_{j-1})},\hspace{5mm}j= (2,..., n)
 \end{eqnarray}
 \begin{eqnarray}
 B_j &=& \sqrt{\frac{\kappa^{\parallel}_j}{\kappa^\perp_j}\xi^2+\frac{i2C_j\rho_j(\omega)}{\kappa^\perp_j}}
 \end{eqnarray}
 
 and,
 
 \begin{equation}
      A_n = -\tanh(B_n d_n)^s 
 \end{equation}
 
 \noindent where $n$ is the number of layers counting from the top surface, i.e., $n=1$ at the surface where the pump and probe lasers are focused, $n=$ ``total number of layers" at the bottom layer, $C_j$ and $\rho_j$ are the specific heat and density of each layer, respectively. The parameter $s$ sets the type of boundary condition at the bottom layer with $s=0$ for a semi-infinite substrate. When the substrate thickness is finite, $s=1$, for adiabatic boundary conditions, and $s=-1$, for the case of isothermal boundary conditions. Finally, the thermal boundary conductance between two layers, $K_{TBC}$, is modeled using the usual assumptions, i.e., a  1 nm thick layer with a small heat capacity (e.g, $C_p \approx$ 1 JKg$^{-1}$C$^{-1}$).

\section{Experimental Details}
\begin{figure}[t]
 \includegraphics[scale=0.65,angle=-0]{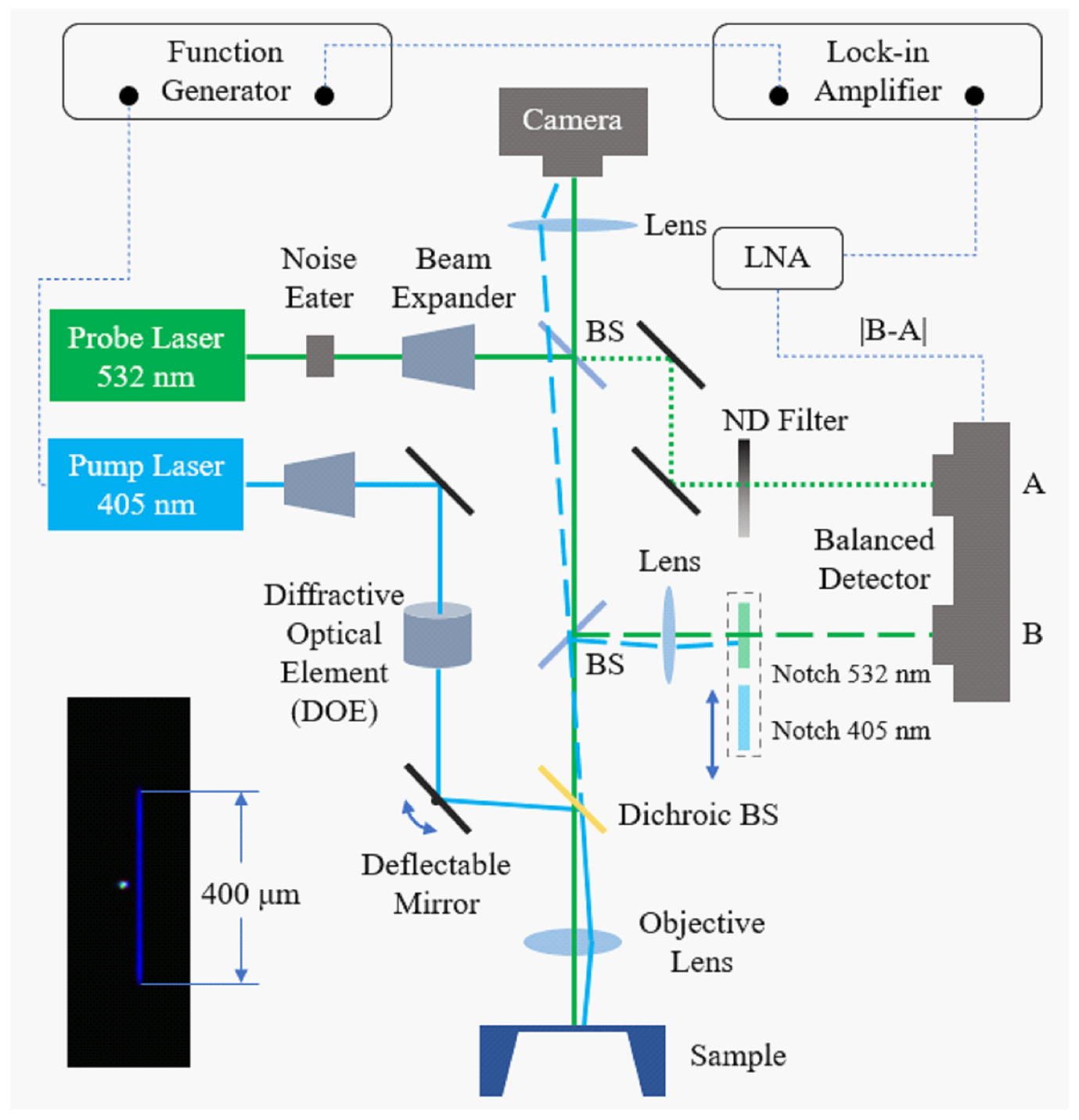}
 \caption{Schematic illustration of the experimental setup used to study in-plane heat transport based on a balanced detection scheme. The inset shows an optical photograph of the line-shaped heat source (pump) as well as the spot-shaped probe. A diffractive optical element (DOE) is used to modify the shape of the pump beam and a deflectabe mirror is used to control the offset between the pump and probe lasers.}
 \label{fig2}
 \end{figure}

Figure \ref{fig2} displays a schematic illustration of the experimental setup. The wavelength of the pump laser was set to $\lambda_{pump}$ = 405 nm (Omicron A350), whereas the probe laser wavelength was set to $\lambda_{probe}$ = 532 nm (Cobolt SAMBA series). The pump laser power was harmonically modulated (intracavity modulation) using a wave function generator (Rigol 54DG5332) between 10 Hz and 100 kHz, whereas the probe laser intensity was approximately constant in time. The pump laser power was continuously increased until a temperature rise of the order of several kelvin was observed (between 1 and tens of mW depending on the thermal properties of the sample), whereas the power of the probe laser was kept low in order to avoid any additional heating effects (typically of the order of tens of µW). A noise eater (Thorlabs NEL01A/M) was used to reduce the intrinsic noise from the probe laser, which was particularly useful for frequencies below 5 kHz. Both lasers were coupled into the main optical axis using beam splitters (BS) and finally focused on the sample through the objective lens (Mitutoyo MY10X-803). The diameters of the pump and probe beams at the output of the lasers were 1100 $\mu$m and 800 $\mu$m, respectively, and their spatial intensity distribution was Gaussian.

\begin{figure*}[t]
 \includegraphics[scale=0.6,angle=0]{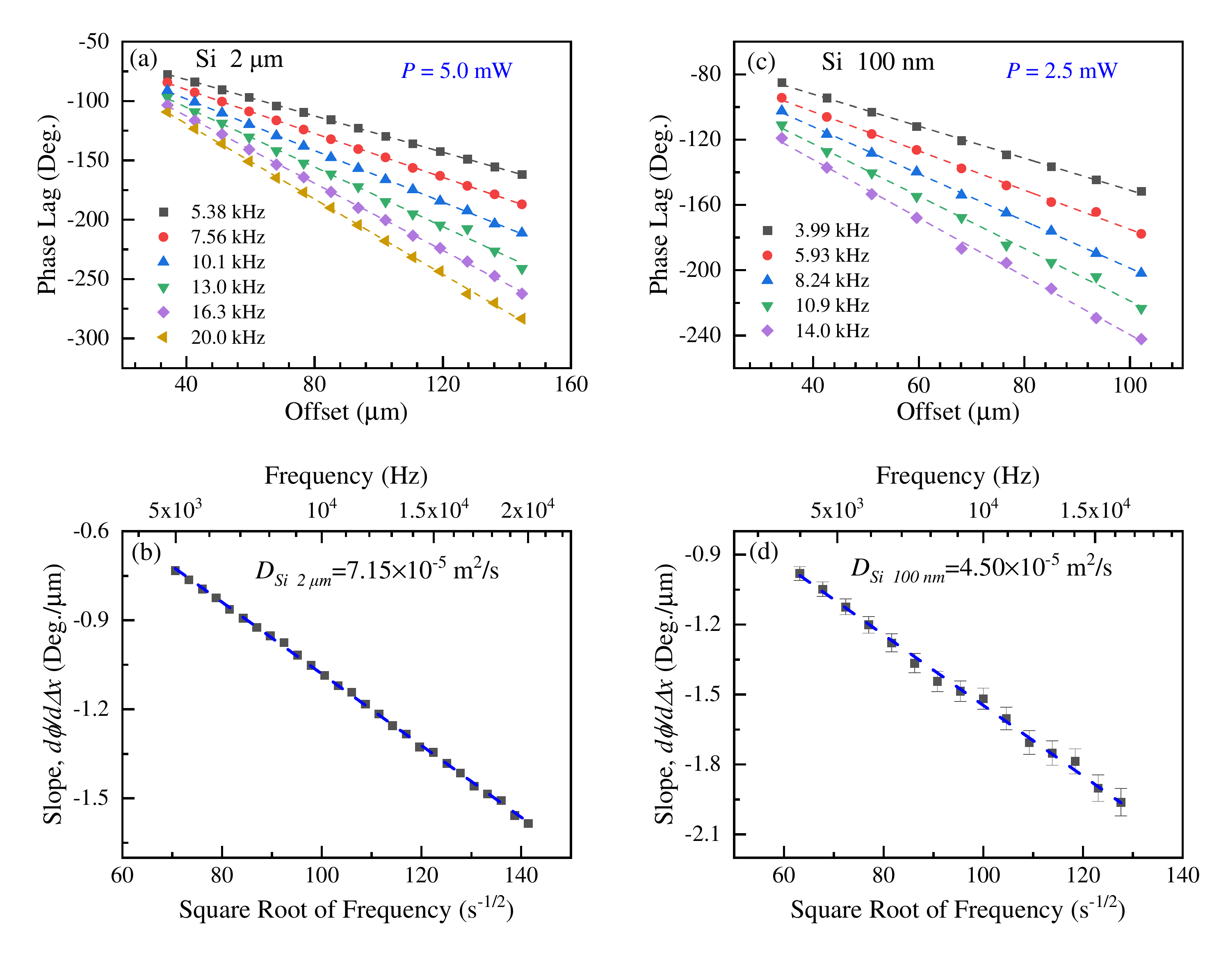}
 \caption{(a) Phase lag vs. offset for a 1 $\mu$m  thick suspended Si thin film for different excitation frequencies between 5 kHz and 20 kHz. The dashed lines are linear fits to the data points. (b) Slopes obtained from fitting $\phi$ vs $\Delta x$ for each excitation frequency. (c) Phase lag vs. offset for a 100 nm  thick suspended Si thin film for different excitation frequencies between 4 kHz and 14 kHz. The dashed lines are linear fits to the data points. (d) Slopes obtained from fitting $\phi$ vs $\Delta x$ for each excitation frequency.}
 \label{fig3}
 \end{figure*}


In order to obtain a line-shaped uniform intensity distribution of the pump beam, we have used a custom designed holographic diffractive optical element (DOE) purchased from HOLO/OR Ltd, which was placed in the optical path of the pump beam. The diameter of the pump beam was tuned to meet the specifications (beam input diameter) of the DOE using a variable optical beam expander. We note that the DOE is one of the key elements in this experimental setup, since it is responsible for homogenizing the intensity of the pump laser, i.e., the incident Gaussian distribution is converted into a line-shaped spot with uniform intensity distribution along the main axis, and with diffraction limited Gaussian distribution in the perpendicular direction after being focused. 
Regarding the probe laser, we simply use a diffraction limited focused Gaussian beam as shown in the schematics of Figure \ref{fig2}. The spatial displacement between the heater (line-shaped pump beam) and the detection (spot-shaped probe beam) was controlled using a deflectable mirror prior to the focusing objective, which was calibrated using a Complementary-Metal-Oxide-Semiconductor (CMOS) camera placed on the object plane of the objective.

\begin{figure}[t]
 \includegraphics[scale=0.6,angle=0]{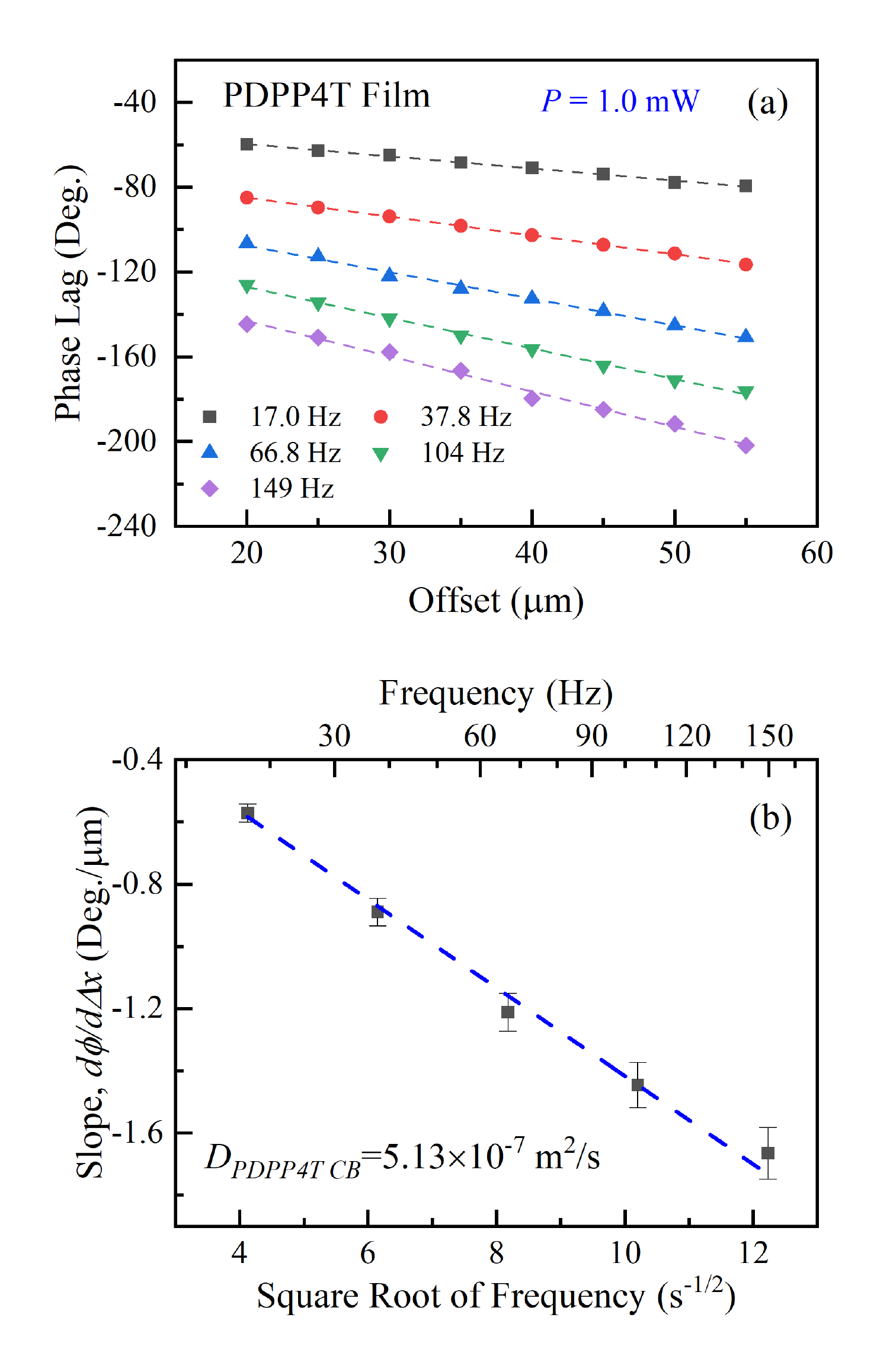}
 \caption{(a) Phase lag vs. offset for a 100 nm  thick PDPP4T suspended thin film for different excitation frequencies between 17 Hz and 150 Hz. The dashed lines are linear fits to the data points. (b) Slopes obtained from fitting $\phi$ vs $\Delta x$ for each excitation frequency.}
 \label{fig3polymer}
 \end{figure}

The back reflection of the pump and probe lasers after focusing on the sample was coupled into the detection arm using a 90:10 BS (R:T), and the laser wavelength was selected by two mechanically controlled notch-filters, and further purified using interferometric filters. As detector we used a large area (5 mm in diameter) AC-coupled balanced detector (Thorlabs PDB210A/M-AC-SP). A sample of the probe laser (532 nm) was independently coupled to the second detector input. The output voltage of the detector is sensitive to the difference between both inputs provided that the difference signal is time-dependent. This is achieved by placing high pass filter before the amplification stage inside the detector (AC-coupled detector). Hence, the output of the balanced detector is a frequency modulated voltage arising from the thermal oscillations on the surface of the sample. In order to minimize the laser noise, the optical path of the signal and reference laser were kept similar within 1 cm, and the input power on both optical detectors was balanced. The pump laser was also coupled to the detection system in order to obtain a reference phase for the thermal signal using the same beam path, i.e. first the pump laser signal (amplitude and phase) was measured by the lock-in amplifier after which the notch-filters were mechanically switched and the amplitude and phase of the thermal signal was obtained. This process was sequentially repeated for each excitation frequency.

\section{Results and Discussion}
We have applied the methodology described in the previous sections to several samples with different heat flow geometries. For all studied samples we have measured the dependence of the phase lag ($\phi$) between the thermal excitation (pump) and the response of the sample (probe) as a function of their spatial offset, and for different excitation frequencies. In particular, we have studied 1D heat flow for two suspended Si thin films with large surface areas (in the cm$^2$ range), and with thicknesses of 2 $\mu$m and 100 nm which we have purchased from NORCADA, as well as a 100 nm thick suspended PDPP4T polymer thin film fabricated in-house.{\color{red} \cite{Ref}} The case of 2D heat flow was addressed for a bulk Bi substrate, a Si bulk substrate covered with 60 nm Au (transducer), a glass bulk substrate covered with 60 nm Au (transducer), and a highly-oriented pyrollitic graphite (HOPG) bulk sample with 60 nm Au (transducer). 


\subsection{Quasi-2D samples without transducer}
Figure \ref{fig3}a displays the phase lag ($\phi$) versus the spatial offset ($\Delta x$) between the pump and probe beams for a suspended Si thin film with a thickness of 2 $\mu$m, and for different thermal excitation frequencies in the range between $\approx$5 kHz and 20 kHz. All suspended thin films where measured in vacuum at a pressure lower than 10$^{-7}$ mbar. In all cases, the data were fitted using a linear relation, as predicted by Eq. \ref{eq3}. The collected data for each frequency consists of multiple data points acquired at different offsets, $\Delta x$, which are uniformly distributed in the range from $\approx$30 $\mu$m to 145 $\mu$m. 
As shown in Figure \ref{fig3}a, the magnitude of $\phi$ increases with increasing frequency and $\Delta x$. 
Figure \ref{fig3}b displays the slopes, $\partial\phi/\partial\Delta x$, as obtained from the linear fits of $\phi$ versus $\Delta x$ in Fig. \ref{fig3}a, at each excitation frequency. In all cases, the experimental errors are within the size of the symbols. 
In good agreement with the prediction of Eq. \ref{eqderivative}, the slope, $\partial \phi/\partial\Delta x$, exhibits a linear relation with $f^{1/2}$.
Hence, using Eq. \ref{eqderivative} we obtained the thermal diffusivity for the 2 $\mu$m thick suspended Si thin film as $D=7.15\times10^{-5}$ m$^2$/s, i.e., 82\% of the Si bulk value and in good agreement with previous determinations.\cite{Chavez-Angel2014}
Figure \ref{fig3}c displays the corresponding data for the 100 nm thick suspended Si thin film. The data exhibits a similar dependence of $\phi$ on $\Delta x$ and $f^{1/2}$, as compared to the 2 $\mu$m thick suspended Si thin film. However, due to the lower thermal diffusivity of the 100 nm thick suspended thin film, we used a slightly lower $\Delta x$ and $f$ ranges.
We note that for this sample the phase lag data points exhibit larger experimental noise, which partly arises from the comparatively lower signal (absolute temperature rise, $\Delta T$) for similar $\Delta x$ and frequency ranges, and which originates from its lower thermal diffusivity. In addition, this sample is more sensitive to additional noise arising from mechanical vibrations introduced by the environment during the measurement process. Using Eq. \ref{eqderivative} we have obtained the thermal diffusivity of the 100 nm thick sample as, $D= (4.50 \pm 0.23) \times10^{-5}m/s^{2}$, i.e., 50\% of the Si bulk value and also in good agreement with previous determinations.\cite{Chavez-Angel2014} The reduced values of the thermal diffusivity obtained for both suspended Si thin films arise from their different thickness, which limits the maximum phonon mean free path of the phonons which can transport heat, a well known effect which was already discussed by Ch\'avez-Angel\cite{Chavez-Angel2014} and references therein.

Figure \ref{fig3polymer} displays the phase lag vs spatial offsets for a suspended PDPP4T thin film with a thickness of 100 nm. The frequency range chosen  was reduced with respect to the case of both Si suspended thin films due to the lower thermal diffusivity of PDPP4T, which leads to a comparatively smaller in-plane thermal penetration depth. In general, the response of this sample is qualitatively similar to the case of both suspended Si thin films, hence, we have applied the same methodology to analyze the experimental data. We have obtained an in-plane thermal diffusivity of $D=(5.13\pm0.26)\times10^{-7} m/s^{2}$, which is a typical value for semi-crystalline polymer thin films.

\subsection{Bulk sample without transducer}

\begin{figure}[t]
 \includegraphics[scale=0.6, angle=0]{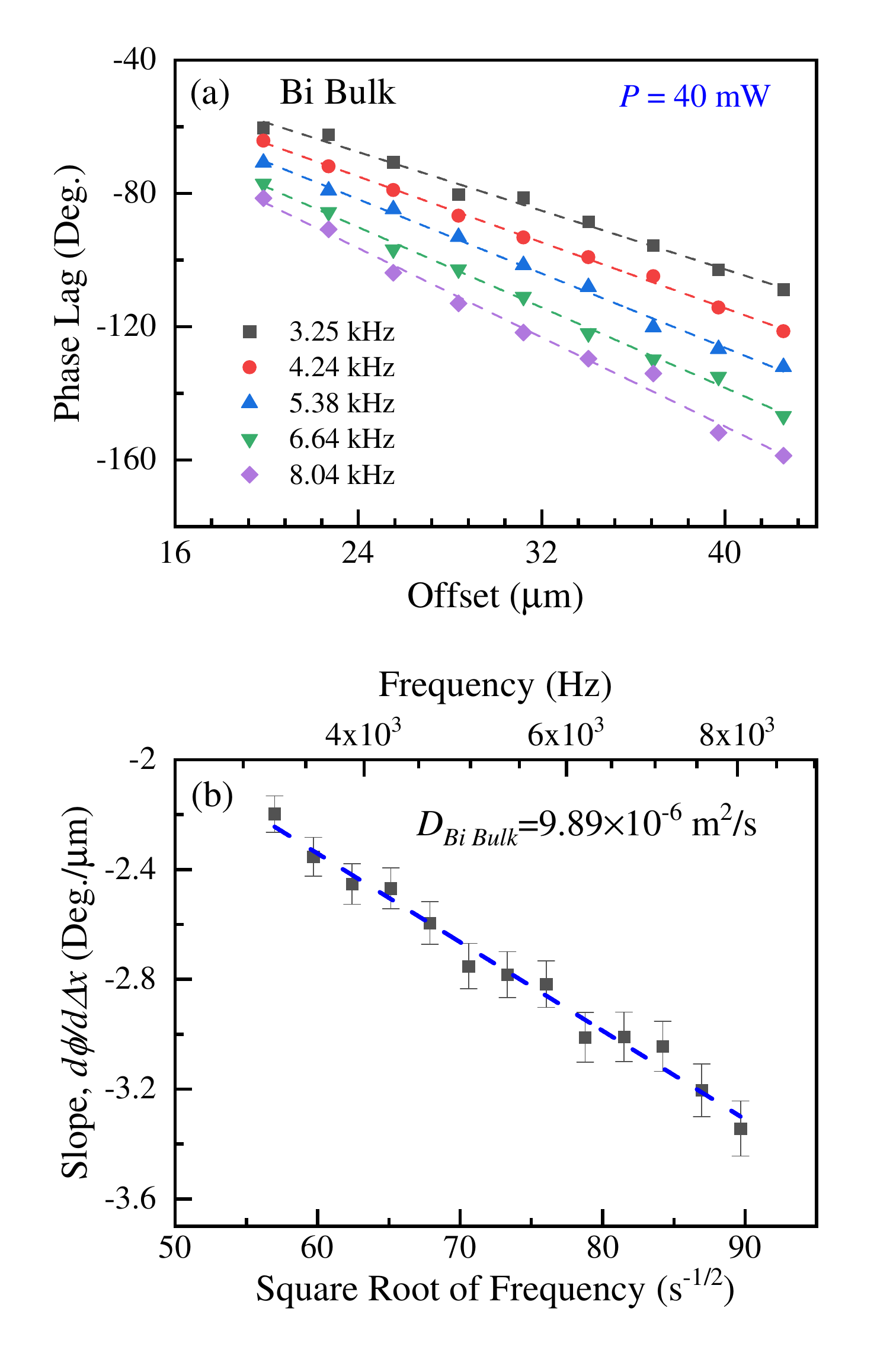}
 \caption{(a) Phase lag vs. offset for a Bi single crystal, and for different excitation frequencies. (b) Slopes obtained for fitting $\phi$ vs $\Delta x$ for each excitation frequency}
 \label{fig4}
 \end{figure}

We have also studied the case of 2D heat flow as discussed in Section \ref{2Dheatflowbulkwotransducer}. For this purpose, we used single crystal Bi(0001) substrate purchased from Mateck GmbH. The pump and probe lasers focused directly onto the surface of the sample, i.e., without using a metallic transducer as for the suspended thin films in the previous section.
Figure \ref{fig4}a displays the phase lag as a function of the spatial offset, and for different excitation frequencies. Similarly as for the case of the suspended thin films studied in the previous section, the dependence of $\phi$ on $\Delta x$ and $f^{1/2}$ is also linear as predicted by Eq. \ref{eqderivative2}. However, unlike the case of the suspended Si thin films, the heat flow geometry for a bulk sample is 2D, i.e., heat propagates in the directions perpendicular to the line-shaped heater. In this case, the spatial dependence of the temperature field is given by Eq. \ref{T2Dheatflow}.
In fact, the phase lag and the corresponding slopes, $\partial \phi /\partial \Delta x$, shown in Fig. \ref{fig4} are prone to larger experimental error as compared to the case of the suspended Si thin films (1D heat flow), which is mainly due to the faster spatial decay of signal as predicted by Eq. \ref{T2Dheatflow}. In addition, the different temperature coefficients of reflectivity (at the probe wavelength of 532 nm) of Bi as compared to Si also account for the different signal levels observed in these systems. The thermal diffusivity of the Bi sample was obtained fitting the slopes ($\partial \phi /\partial \Delta x$) shown in Fig. \ref{fig4}b, obtaining $D=9.89\times10^{-6} m^2/s$, which is in good agreement with previously reported values.\cite{Gallo2004}

\subsection{Bulk samples with transducer}
\begin{figure*}[t]
 \includegraphics[scale=0.6,angle=0]{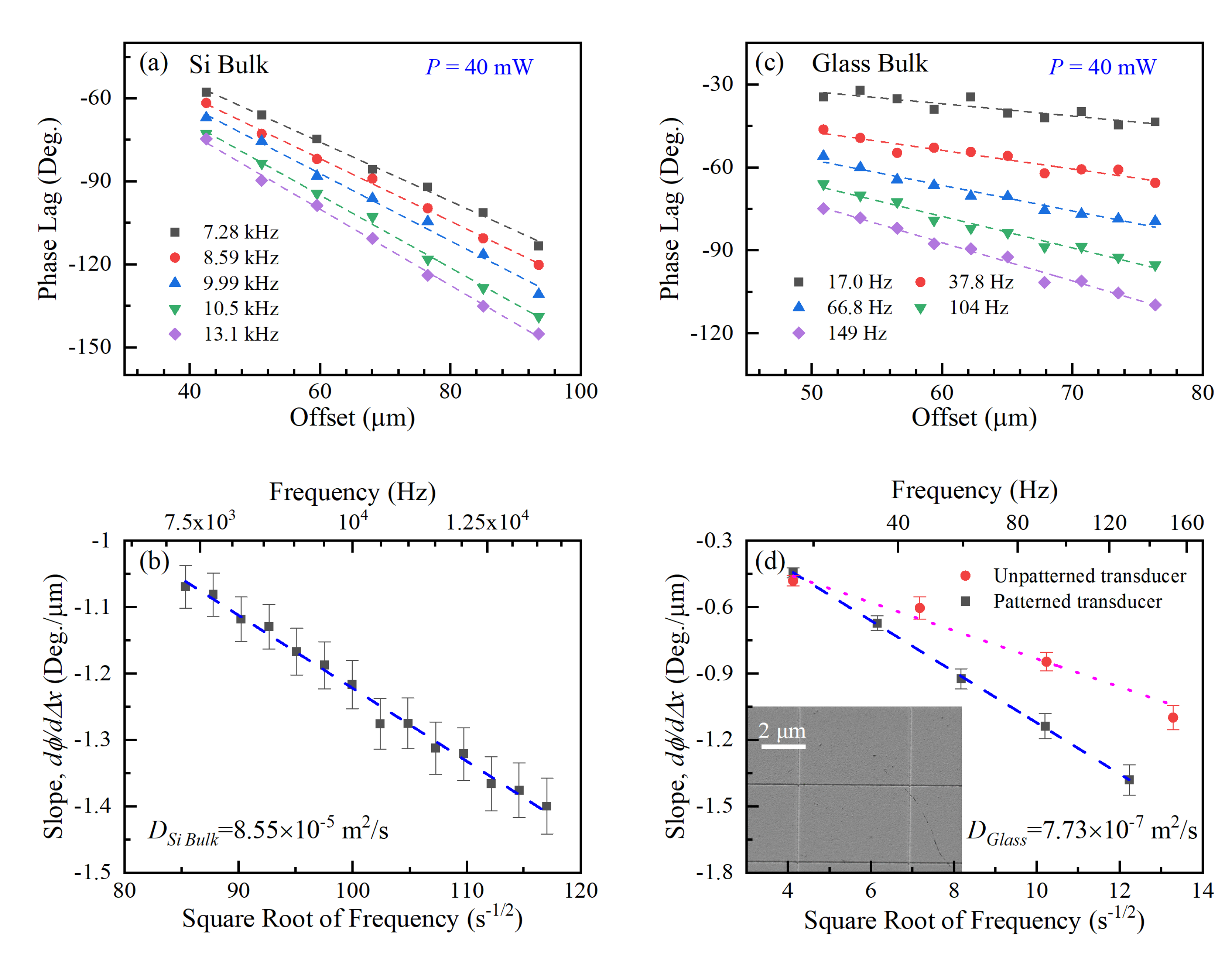}
 \caption{(a) Phase lag vs. offset for a Si substrate with a 60 nm thick Au transducer for different excitation frequencies between 7 kHz and 14 kHz. The dashed lines are linear fits to the data points. (b) Slopes obtained from fitting $\phi$ vs $\Delta x$ for each excitation frequency. (c) Phase lag vs. offset for a glass substrate with a 60 nm thick patterned Au transducer for different excitation frequencies between 17 Hz and 150 Hz. The dashed lines are linear fits to the data points. (d) Slopes obtained from fitting $\phi$ vs $\Delta x$ for each excitation frequency for the as-deposited Au transducer as well as for the patterned transducer. The inset displays and image of the pattern used to avoid in-plane heat transport within the Au transducer.}
 \label{fig5}
 \end{figure*}

We have also studied the influence of using a Au metallic transducer on the phase lag response.
The use of a metallic transducer in thermoreflectance experiments is usually desirable for three main reasons: (i) the extinction coefficient of a metal for visible incident light is typically rather small ($<$100 nm), hence, allowing to approximate the heat source as superficial; (ii) no direct excitation of the electronic systems of the material under investigation occurs, avoiding heat transport through excitons which could be relevant for some materials with indirect optical bandgap; and (iii) the temperature coefficient of reflectance of Au is relatively large, $\Delta R/R_0 \approx 2\times 10^{-4}$, for 532 nm probe wavelength, which naturally results in enhanced thermal signal.
In particular, we address to what extent the present method can be applied to multilayered systems. Typically, the presence of a metallic transducer can have a strong impact on the measured phase lag, which originates from in-plane heat conduction in the transducer layer itself, and which is particularly relevant when the thermal conductivity of the transducer ($\kappa_{tr}$) is much larger than the thermal conductivity of the substrate ($\kappa_s$). We have chosen as case study a bulk crystalline Si substrate as well as a bulk glass substrate, hence, addressing the cases with $\kappa_{tr} \approx \kappa_{s}$ (Si), and $\kappa_{tr} \gg \kappa_{s}$ (glass). For the case with $\kappa_{tr} \ll \kappa_{s}$, the signal is fully dominated by the substrate, as shown numerically solving Eq. \ref{solution_multilayers} in the Supporting Information section.

Figure \ref{fig5}a displays the phase lag data for a Si substrate covered with a 60 nm thick Au transducer, which was deposited using thermal evaporation with a base pressure better than 10$^{-6}$ mbar, and at a deposition rate of 1 $\AA/s$. The phase lag curves were collected at excitation frequencies between 7 kHz and 13 kHz, and spatial offsets between 40 $\mu$m and 90 $\mu$m. The data exhibits a similar trend as observed for the case of bulk Bi, i.e., even in the presence of the Au transducer the phase lag depends linearly on the spatial offset. Figure \ref{fig5}b displays the slopes as extracted from the linear fits of the phase lag curves in Fig. \ref{fig5}a as a function of $f^{1/2}$. Once more, the relation between $\partial \phi / \partial \Delta x$ and $f^{1/2}$ is linear. We applied Eq. \ref{eqderivative2} to compute the thermal diffusivity of the sample obtaining $D=(8.55\pm 0.43)\times10^{-5} m^2/s$, which is in good agreement with reported values for Si substrates.\cite{Shanks963,Glassbrenner1964,Fulkerson1968}

It is interesting to note that even in the presence of the Au transducer, it is still possible to directly obtain the thermal diffusivity of the substrate from the double derivative of the phase lag, $\partial^2\phi/[\partial{\Delta x} \partial{f^{1/2}}]$, as previously done for the case of the suspended thin films and for the Bi substrate without transducer. We have confirmed this approximation solving numerically Eq. \ref{solution_multilayers} for the studied geometry, and observing that when the thermal conductivity of the transducer is similar or lower to the thermal conductivity of the underlying substrate, the response is mostly dominated by the substrate and, hence, the present method can be applied without significant loss of accuracy, which is estimated in $ < 2\%$ for the present case. 

\begin{figure}[t]
 \includegraphics[scale=0.6,angle=0]{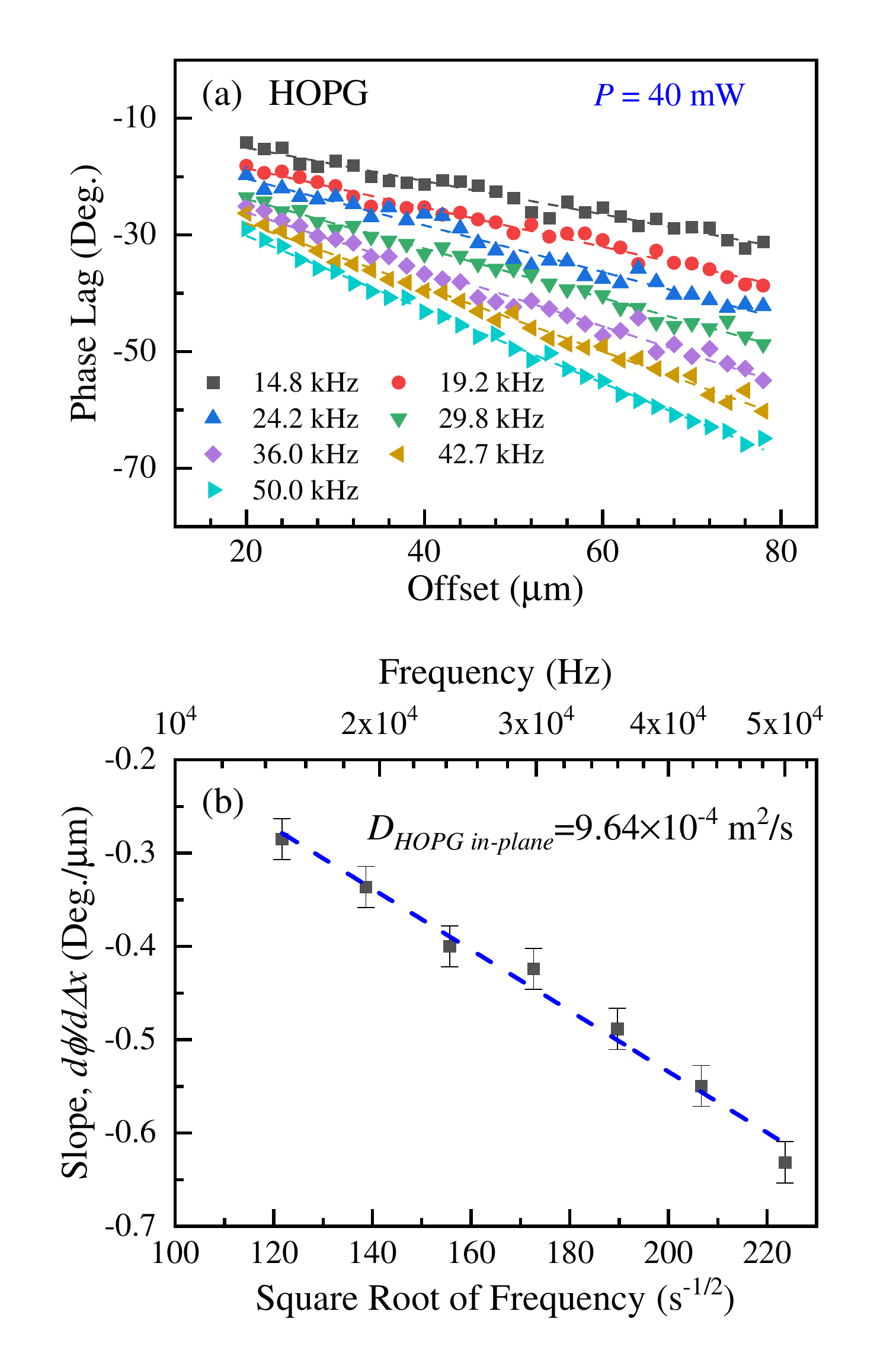}
 \caption{(a) Phase lag vs. offset for a HOPG bulk sample with a 60 nm thick Au transducer for different excitation frequencies between 14 kHz and 50 kHz. The dashed lines are linear fits to the data points. (b) Slopes obtained from fitting $\phi$ vs $\Delta x$ for each excitation frequency.}
 \label{fig7}
 \end{figure}

We also discuss the case of a glass substrate covered with a 60 nm thick Au transducer deposited through thermal evaporation in similar conditions as for the case of the Si substrate. The thermal conductivity of the Au transducer was determined by measuring its electrical conductivity through the Van der Pauw method, and using the Wiedemann-Franz law obtaining, $\kappa_{tr}=246$ Wm$^{-1}$K$^{-1}$. This case is substantially different from the previous since the thermal conductivity of the Au transducer is more than two orders of magnitude larger than the thermal conductivity of the glass sample ($\kappa_{Glass} \approx 1$). In fact, it is already expected that the response will be dominated by a combination of both, the glass sample and the Au transducer. In order to further study this effect, we have patterned the Au transducer using focused ion beam with the purpose of substantially reducing its in-plane thermal conductivity. Hence, we provide experimental data for the case of a glass substrate with a continuous Au transducer, $\kappa_{//}\approx \kappa_{\perp}$, as well as for the case with a patterned Au transducer, $\kappa_{//} \approx 0$. The patterned Au transducer was fabricated in a 150 $\mu m$ $\times$ 150 $\mu m$ area, and was obtained patterning narrow ($\approx$ 15 nm) vertical and horizontal channels on the as-deposited Au transducer. The depth of these channels was carefully calibrated in order not to damage the surface of the sample, hence, the residual thickness of the transducer at the bottom of each channel is $<$ 3 nm. The horizontal and vertical spacing between each channel was 5 $\mu m$. The inset of Figure \ref{fig5}d displays an optical image of the pattern.
Figure \ref{fig5}c displays the phase lag data for the case of the glass substrate with the patterned Au transducer in the frequency range from 17 Hz to 150 Hz. Note that we have selected a rather low frequency range in order to enhance the thermal penetration depth, $\mu = \sqrt{D/(\pi f)}$, arising from the low thermal diffusivity of the glass substrate and, hence, to increase the relative sensitivity of the thermal properties of the glass substrate with respect to the Au transducer. Figure \ref{fig5}d displays the slopes, $\partial \phi/\partial \Delta x$, as a function of $f^{1/2}$ for both glass samples, i.e., with and without patterned transducer. Fitting the data corresponding to the sample with patterned substrate renders a thermal diffusivity $D=(7.73\pm0.39)\times10^{-7} m^2/s$, which is in good agreement with previously reported values for similar substrates.\cite{Yang1992} In the case of the sample with non-patterned transducer a direct fit to the data points in Fig. \ref{fig5}d renders $D=(2.55\pm0.13)\times10^{-6} m^2/s$, which is obviously larger than the typical values observed for  glass substrates. As we have previously anticipated, the origin of this larger value is heat conduction through the Au transducer. In other words, when in-plane heat transport within the transducer is not negligible, the present method renders the effective thermal conductivity of the system. 

We note that even for the cases where the transducer has a non-negligible contribution to the in-plane thermal conductance of the multilayered system, the response $\phi(\Delta x)$ is still linear for each excitation frequency. However, the second order derivative, $\partial^2\phi/\partial{\Delta x} \partial{f^{1/2}}$, provides the thermal diffusivity of the multilayered systems, which depends on the used experimental conditions such as the offset and excitation frequency ranges. This result is simply a consequence of the expansion of $K_0(qr)$ for $qr \gg 1$. In this case, it is still possible to obtain the thermal diffusivity of the sample through numerical simulations reflecting the experimental conditions.

Finally, in Fig. \ref{fig7} we show the phase lag data for a HOPG bulk sample purchased from SPI Supplies. This sample was studied using a 60 nm thick Au transducer in order to enhance the acquired thermal signal. We have also studied this sample in the absence of the metallic transducer, however, almost not signal could be obtained due to its low temperature coefficient of reflectance for the used pump and probe wavelengths. We note that the presence of the metallic transducer has a negligible effect on the response, similarly as for the case of the Si substrate. In fact, the HOPG substrate exhibits a rather large in-plane thermal diffusivity and, hence, it is insensitive to large extent to the effect of the transducer. We have estimated a deviation below 1\% numerically solving Eq. \ref{solution_multilayers}.
Figure \ref{fig7}b displays the slopes, $\partial \phi/\partial \Delta x$, as a function of $f^{1/2}$ for the HOPG sample with transducer. Fitting the experimental data renders a thermal diffusivity $D=(9.64\pm0.48)\times10^{-4} m^2/s$, which is a typical value for the in-plane thermal diffusivity this system.\cite{Qian2020} Interestingly, the out-of-plane thermal diffusivity of this sample is typically $\approx$150 times smaller than its in-plane counterpart, which evidences the potential of the present approach to study in-plane heat transport for samples with large thermal anisotropy ratios (in-plane vs. out-of-plane).

\section{conclusions}
We have developed a new method based on beam-offset frequency-domain thermoreflectance (BO-FDTR) suitable to study thermal transport with enhanced sensitivity to in-plane heat flow. The present method substantially differs from previous approaches since it uses a one-dimensional (1D) heat source combined with controlled spatial offsets between the heater (pump, line-shaped) and the thermometer (probe, dot-shaped). The key advantage of using a 1D heat source with respect to focused Gaussian heat sources (0D) relies on the spatial dependence of the temperature field. Using a 1D heat source allows to probe the thermal field at larger distances from the heat source, providing enhanced sensitivity to in-plane thermal transport. In addition, the spatial dependence of the thermal response at distances larger than the typical dimensions of the heat source is insensitive to the specific shape of the heat source. Regarding the data analysis procedure, this approach presents large advantages with respect to other methods since it does not require any computational efforts, i.e., the thermal diffusivity of the specimen is readily extracted from linear fits to the data points. In addition, the use of a line-shaped heater defines the frequency region of interest to frequencies $<$ 100 kHz, which provides substantial advantages regarding the price of the excitation source. Most importantly, such low frequency range allows to study samples without the use of a metallic transducer minimizing the influence of the finite optical penetration depth in the samples due to the large thermal penetration depth provided by the low frequency range.

We have applied the present method to study 2D and 3D samples without the use of a metallic transducer, showing that the thermal diffusivity can be accurately extracted for both sample geometries. Furthermore, we show that the method is still accurate for samples with a thin metallic transducer, provided that in-plane heat transport within the transducer is negligible. In other words, when the in-plane component of the thermal diffusivity of the transducer is similar or smaller than the thermal diffusivity of the sample, the spatial dependence of the thermal field is mostly dominated by the sample. For cases when the in-plane component of the thermal diffusivity of the transducer cannot be neglected, we developed a strategy based on patterning the metallic transducer using focused ion beam by patterning channels on the transducer which have the effect of substantially reducing the in-plane component of the thermal diffusivity of the transducer.

We think that the present method will provide new opportunities to study heat transport, specially for anisotropic materials, since it allows to obtain the in-plane components of the thermal conductivity/diffusivity tensor with enhanced sensitivity. Furthermore, its rather simple data analysis procedure makes it suitable for cases where numerical simulations cannot be conducted

\section{Acknowledgements}
We acknowledge financial support from the Spanish Ministerio de Econom\'ia, Industria y Competitividad for its support through grant CEX2019-000917-S (FUNFUTURE) in the framework of the Spanish Severo Ochoa Centre of Excellence program, and grants PID2020-119777GB-I00 (THERM2MAIN), and PDC2021-121814-I00 (COVEQ). This work was financially supported by the European Commission through the Marie Sklodowska-Curie project HORATES (GA-955837). We also acknowledge grant 2021-SGR-00444 (NANOPTO) from AGAUR. K.X. acknowledges a fellowship (CSC201806950006) from China Scholarship Council and the PhD programme in Materials Science from Universitat Autònoma de Barcelona in which he was enrolled. G.R and I.Z. acknowledge financial support from Eucor, The European Campus (Marie Sklodowska-Curie QUSTEC grant agreement no. 847471) and the Swiss National Science Foundation (Project Grant No. CRSII5 189924).

\section{Data Availability Statement}
All data needed to evaluate the conclusions in the paper are present in the paper. Additional data related to this paper may be requested from the authors.

%

\end{document}